\documentclass[12pt]{article}

\usepackage{newtxtext,newtxmath}

\usepackage{graphicx}

\usepackage[letterpaper,margin=1in]{geometry}

\renewenvironment{abstract}
	{\quotation}
	{\endquotation}

\date{}

\makeatletter
\renewcommand{\fnum@figure}{\textbf{Figure \thefigure}}
\renewcommand{\fnum@table}{\textbf{Table \thetable}}
\makeatother

\usepackage{scicite}

\usepackage{url}

\usepackage{subcaption}
\usepackage{hyperref}

\def\scititle{
    Freshness, Persistence and Success \\ of Scientific Teams
}
\title{\bfseries \boldmath \scititle}

\author{
	Hanjo D. Boekhout$^{1\ast}$,
	Eelke M. Heemskerk$^{2}$,
	Niccol\`{o} Pisani$^{3}$,
    Frank W. Takes$^{1}$\and
	\small$^{1}$Department of Computer Science (LIACS), Leiden University, The Netherlands.\and
	\small$^{2}$Department of Political Science, University of Amsterdam, The Netherlands.\and
    \small$^{3}$IMD Business School, Lausanne, Switzerland.\and
	\small$^\ast$Corresponding author. Email: h.d.boekhout@liacs.leidenuniv.nl
}

\begin{document} 

\maketitle

\begin{abstract} \bfseries \boldmath
Team science dominates scientific knowledge production, but what makes academic teams successful? 
Using temporal data on 25.2 million publications and 31.8 million authors, we propose a novel network-driven approach to identify and study the success of persistent teams. 
Challenging the idea that persistence alone drives success, we find that team \emph{freshness} --- new collaborations built on prior experience --- is key to success. 
High impact research tends to emerge early in a team’s lifespan. 
Analyzing complex team overlap, we find that teams open to  new collaborative ties consistently produce better science. 
Specifically, team re-combinations that introduce new freshness impulses \emph{sustain} success, while persistence impulses from experienced teams are linked to \emph{earlier} impact. 
Together, freshness and persistence shape team success across collaboration stages.

\end{abstract}

\newpage

\noindent
Team science has become the dominant form for producing academic knowledge across disciplines, generating more highly cited and high-impact research than scholars working independently~\cite{wuchty2007increasing}. 
It is well-known that researchers who collaborate with others are far more likely to produce a ‘hit’ paper in the top 5\% of citations~\cite{mukherjee2017nearly}. 
Teams are successful because they effectively facilitate the recombination of novel ideas with existing knowledge~\cite{uzzi2013atypical}. 
This ability to produce relevant knowledge outweighs the significant coordination costs team science imposes on its members to bridge institutional differences and geographic distance, as long as team size and diversity remain manageable~\cite{cummings2005collaborative,cummings2007coordination,lariviere2015team,hsiehchen2015multinational}. 
Still, some teams are better able to produce relevant knowledge than others. 
We present a novel approach to explain why.

The team performance literature convincingly and consistently argues that team success is crucially dependent on well-balanced intergroup dynamics and leadership practices~\cite{tannenbaum2020teams}. 
When teams invest in getting their team dynamics right, their collaborative learning increases success.  The benefit of this type of \emph{persistent} collaboration was found to be greater for heterogeneous teams, with coauthors having diverse scientific impact, disciplinary background, or scientific ages, than for homogeneous compositions, even as they face higher coordination costs~\cite{bu2018understanding}. Consistent with this colloquial understanding of the benefits of collaboration in science, many of the largest research grant schemes fund consortia of scholars and institutions with a successful track record of collaboration~\cite{wanzenbock2020proposal}. Team success then increases future success chances, consistent with the canonical Matthew effect in science~\cite{merton1968matthew,bol2018matthew}. 

However, another line of research shows that fresh teams with fewer previous collaborations show higher scientific originality and hence perform better because \emph{freshness} crucially fosters knowledge recombination~\cite{zeng2021fresh}. 
For larger teams, evidence suggests that the fresher the better~\cite{liu2022team}. 
But clearly, team freshness is at odds with persistent collaboration. 
This presents us with a paradox: What drives the success of teams in science, persistence or freshness? 

Our starting point is that it is freshness, not persistence, that drives teams to produce high-impact work. 
And that when they are successful, these teams will have a high propensity to keep working together because they want to repeat the success. This cognitive bias is known as the hot-hand effect, where team members believe that after a series of successes, their team is more likely to continue the success~\cite{ayton2004hot}. 
In science, this hot-hand effect has been partially institutionalized by a system-level feedback loop as successful teams have a higher propensity to receive academic grants and awards~\cite{bol2018matthew}.  Team persistence is then primarily a consequence of success, rather than the cause. 
The empirical implication of this is that, in general, scientific success should come sooner rather than later in persistent team collaborations. We therefore engage in an encompassing study of persistent scientific teams and their success. 

Defining what a scientific team actually is, remains far from straightforward. 
Traditional definitions, such as all authors on a publication, all researchers on a grant, or all members of an institute, fail to capture the fluid and interdisciplinary nature of modern science. 
Moreover, it assumes that a team coincides with a piece of output, while we start from the assumption that teams can have several pieces of output.
Moreover, these definitions often impose artificial boundaries, e.g., based on scientific fields or geographical delineations, limiting existing studies to only fragments of the complete scientific landscape. 
To overcome these challenges, we model the dynamics of global science as an evolving scientific co-authorship network, encompassing 25.2 million publications and 31.8 million authors over a span of 13 years. 
Our novel highly scalable network-driven approach first processes this co-authorship network into a persistent collaboration network, where authors are connected for particular periods of time during which their collaborations are sufficiently persistent. 
Persistent teams are then identified as temporal maximal cliques in this network, i.e., as fully connected subsets of authors over an uninterrupted timespan, such that the timespan of the team cannot be further extended while remaining fully connected, nor the team can be extended with an additional member for the same timespan while remaining fully connected throughout the timespan.  
In our study, we leverage the fact that finding these so-called maximal cliques in temporal networks with millions of nodes has recently become possible as a result of algorithmic advances~\cite{boekhout2024fastmaximalcliqueenumeration}.  Note that our first observation of a collaboration is when the involved authors first publish together.  
With this, we can now identify all scientific teams across the globe, without artificial boundaries or selection biases, and study their success. 

Our network approach has the added benefit that it allows us to study more complex properties of scientific collaboration. While teams are typically considered closed systems in empirical studies, in reality teams overlap~\cite{simmel1908soziologie,liu2022team}.  Indeed, teams are part of a complex dynamic scientific ecosystem.  Throughout a team's lifespan, team members can also collaborate with other non-team members, for example, through an incidental extension of the team or when a substantial subset of the team publishes work with others. 
This \emph{openness} to other collaborative work increases the ability of teams to create unusual re-combinations of knowledge crucial for scientific success~\cite{uzzi2013atypical}.  But it can also benefit new teams by building on ongoing successful collaborations. We are particularly keen to uncover when, during their lifespan, teams benefit from certain kinds of collaborations.  

To study this, we need a satisfactory indicator of  \emph{success}. While many definitions exist and come with their obvious benefits and drawbacks~\cite{bai2020quantifying}, we choose a classical definition. 
Each team has a duration and a set of associated timestamped publications used to determine the team's success. 
We deem a publication successful if, within its field and for its year of publication, it belongs to the top 1\% (or top 10\%) most highly cited papers over a period of 3 years after the moment of publication.  
Further methodological details are provided in the Supplementary section Materials and Methods. 
Data for this study is sourced from the in-house version of the Web of Science (WoS) bibliographic dataset at the Center of Science and Technology Studies (CWTS) at Leiden University. 

Altogether, the above gives us the ingredients to uncover, at scale, how team persistence, composition, openness and freshness play a role in understanding the success of scientific teams. 

\section*{Results}
\subsection*{Success of persistent teams}

\begin{figure}[ht]
    \centering
    \begin{subfigure}[t]{0.35\linewidth}
        \centering
        \includegraphics[width=\linewidth]{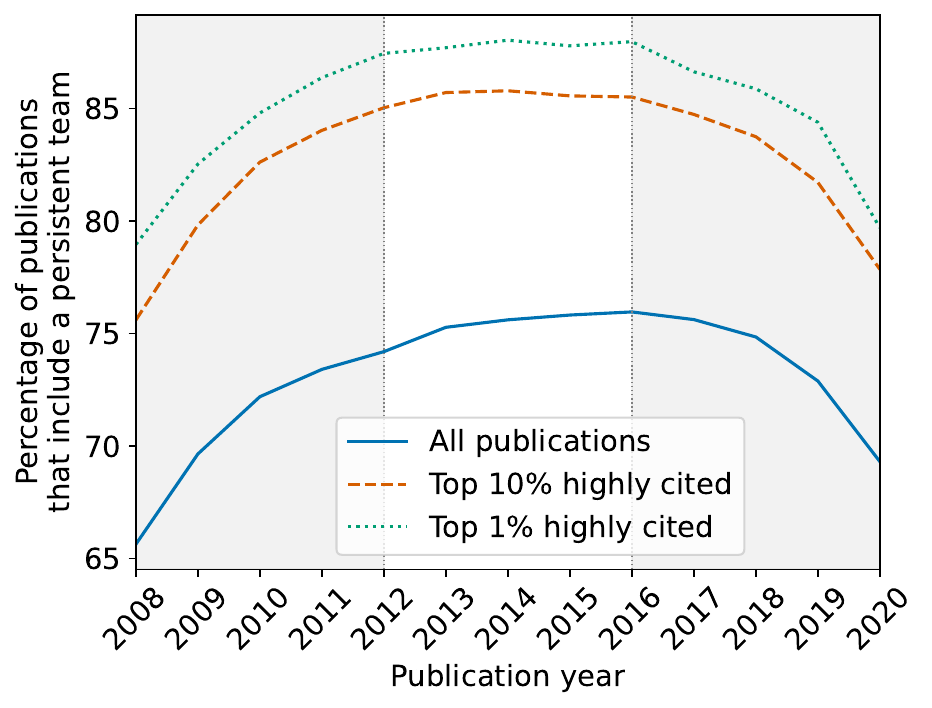}
        \caption{Global team prevalence}
        \label{fig:1a}
    \end{subfigure}
    ~
    \begin{subfigure}[t]{0.63\linewidth}
        \centering
        \includegraphics[width=\linewidth]{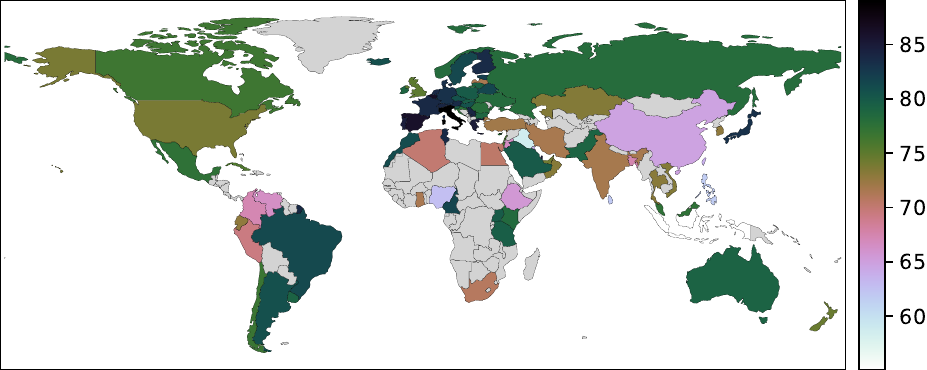}
        \caption{Team prevalence within countries (all publications)}
        \label{fig:1b}
    \end{subfigure}
    \caption{\textbf{Persistent teams among publication author teams (single author teams excluded).} Panel (a) shows the percentage of publications that has at least half of a persistent team's members among their author team globally, for both all and highly cited publications, over time. Panel (b) indicates this percentage aggregated for the set of all publications associated with a specific country.}
    \label{fig:1}
\end{figure}

Figure~\ref{fig:1} confirms that team science is indeed omnipresent today. 
Our network-driven approach identifies a total of 10,232,084 scientific teams with an average duration of 4.37 years. Note that the shoulders of the curve before 2012 and after 2016 are a mere artifact of the choices made for being ``sufficiently persistent''. 
Teams are responsible for over 70\% of global scientific output, and in even greater proportions in highly cited work. Teams are not only the most common container for knowledge creation, their work is also more successful than non-team science (see Figure~\ref{fig:1a}). Of course, some countries are more invested in team science than others. For some European countries team science covers over 85\% of their academic output (see Figure~\ref{fig:1b}).  Noticeably, the USA has a distinctly lower involvement of persistent teams than most Western countries; and upcoming scientific giant China~\cite{pisani2024china} shows one of the lowest rates of involvement of persistent collaborations.  
In what follows, we focus on 1\% highly cited work, with robustness checks for 10\% given in the Supplementary Material. 

\subsection*{Team freshness}

\begin{figure}[!t]
    \centering
    \begin{subfigure}[t]{0.48\linewidth}
        \centering
        \includegraphics[width=\linewidth]{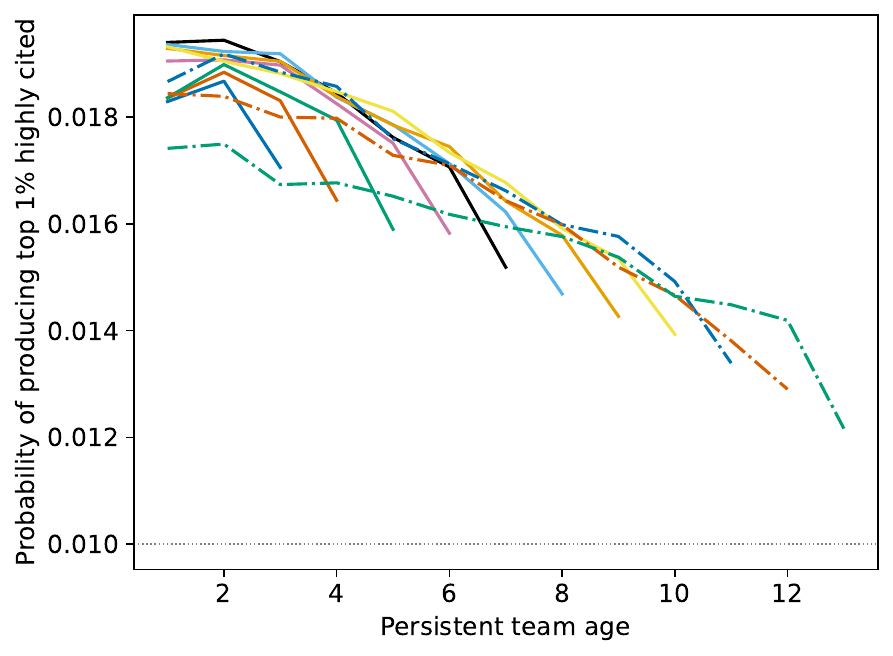}
        \caption{Success probability over time}
        \label{fig:2a}
    \end{subfigure}
    ~
    \begin{subfigure}[t]{0.48\linewidth}
        \centering
        \includegraphics[width=\linewidth]{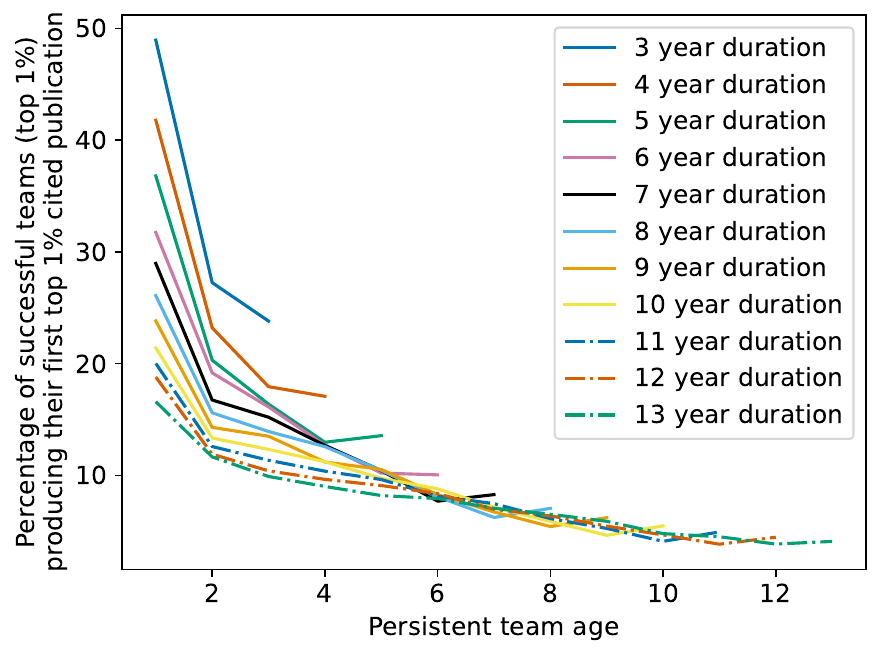}
        \caption{Year of first success}
        \label{fig:2b}
    \end{subfigure}
    \caption{\textbf{Team freshness vs. top 1\% citation success.} Panel (a) shows the probability that publications, produced by teams of a given age, become highly cited. Panel (b) indicates the percentage of highly cited teams 
    that have their first highly cited publication at a given team age.}
    \label{fig:2}
\end{figure}

What then is the relationship between team persistence, freshness, and success in science?  Figure~\ref{fig:2a} reveals that, regardless of the duration of the persistent team, the odds of a successful publication continuously decrease after the second year. This speaks against the idea that persistent collaboration aids the production of highly cited work and instead confirms our intuition that persistent co-authorship follows from a successful fresh collaboration. 
Figure~\ref{fig:2b} shows that a significant portion of persistent teams that become successful actually publish a highly cited paper in their first year.
Furthermore, Figure~\ref{fig:s2addendum} shows that the percentage of not yet successful teams that become newly successful, decreases over time.
Nevertheless, we do find that some teams do not achieve success until their final year and that even in the last year the odds of citation success for persistent teams are higher than average (i.e., higher than the 10\% and 1\% used to define success in the first place). This confirms the implication from Figure~\ref{fig:1} that team science outperforms individual scientific work. However, high-impact work typically comes early on in the collaboration. 

\subsection*{Team composition}

\begin{figure}[!t]
    \centering
    \includegraphics[width=.72\linewidth]{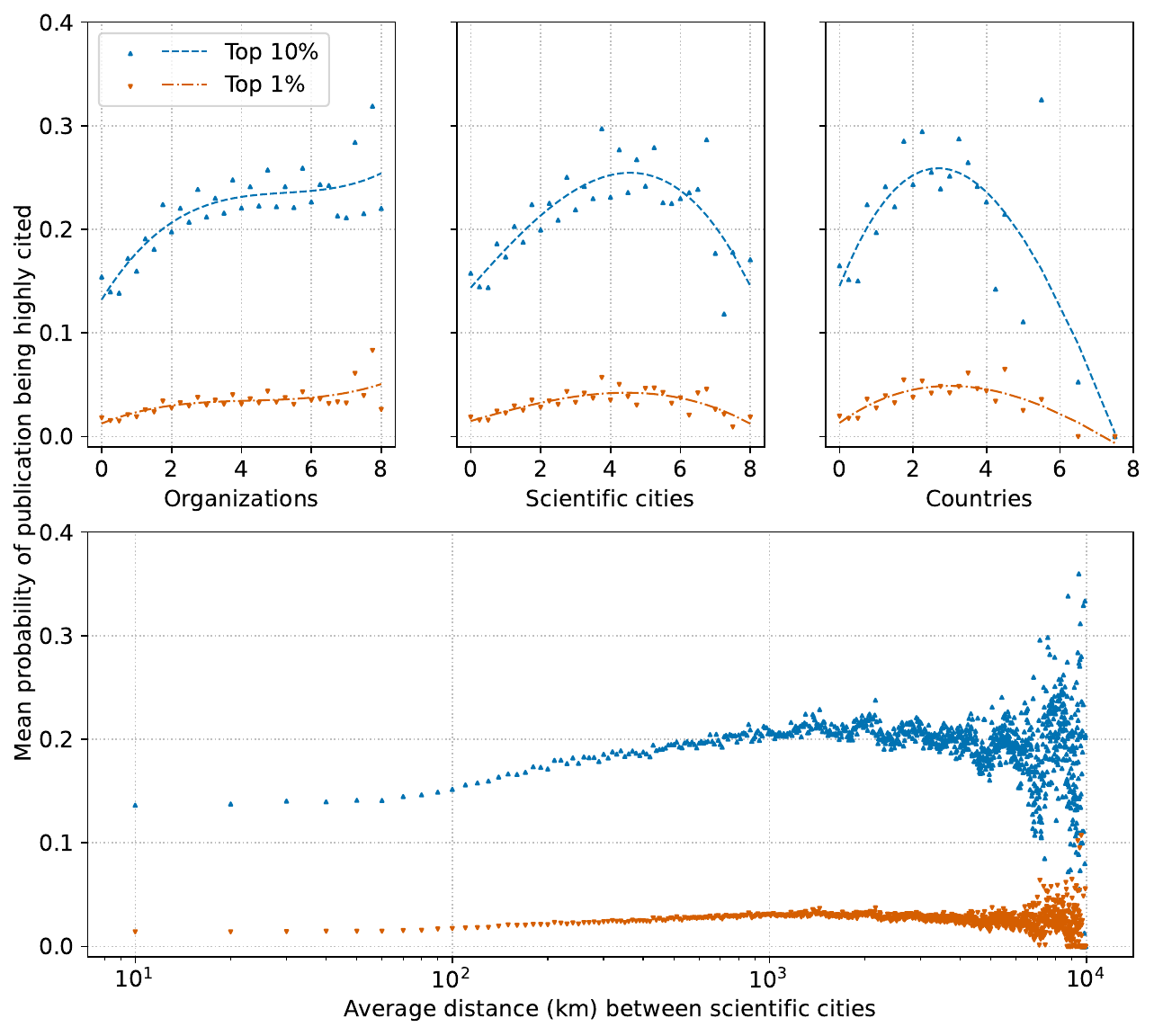}
    \caption{\textbf{Team composition vs. citation success.} The top row panels show the mean probability of citation success for a team's publications, based on the number of organizations, scientific cities, and countries that the team members are associated with on their publications per team member. Results are binned by rounding down to increments of 0.25. The bottom panel shows the same but then for the average distance between scientific cities in kilometers, binned to increments of 10~km.} 
    \label{fig:3}
\end{figure}

How is team success related to team composition? Previous work showed that large and diverse teams, spanning multiple institutions and countries, achieve higher impact~\cite{lariviere2015team}, but that the added benefit per new author diminishes in larger teams. Having more unique countries in a team was however found to consistently increase success~\cite{hsiehchen2015multinational}. To assess this at scale, we computed for each team the average number of distinct locations a team is associated with. After grouping teams in bins based on their locality, we plotted the locality against the average probability that publications of the binned teams will be highly cited (see Figure~\ref{fig:3}). 

We find that teams are more likely to produce highly cited work when they associate with a diverse set of  organizations, cities, and countries. This speaks to the quality of teams to facilitate the recombination of novel ideas with existing knowledge. But, more is certainly not always better. While the probability for success levels out after two organizations are involved per team member, we see decreasing probabilities for teams associated with more than three countries and more than four cities per member. This implies that when institutional differences and geographic distance increase, for example due to their members' mobility, teams are faced with increased coordination costs. As a result, we see that average distances between associated cities greater than a few hundred kilometers, when normalized by team size, provide no further gain in success odds. 

\subsection*{Team openness and success}
So far, we have assumed teams are closed systems and found that freshness rather than persistence is a driving force behind the success of scientific teams. We also observed that, on average, teams have higher odds to produce highly cited work throughout their duration. We can better understand this finding by applying an open system approach that takes into account distinct types of team overlap. While working together in a  team, its members may very well participate in other teams as well. Below, we investigate how teams use this kind of openness through collaborations as a way to strengthen their freshness as well as their persistence.

\begin{figure}[!b]
    \centering
    \includegraphics[width=.9\linewidth]{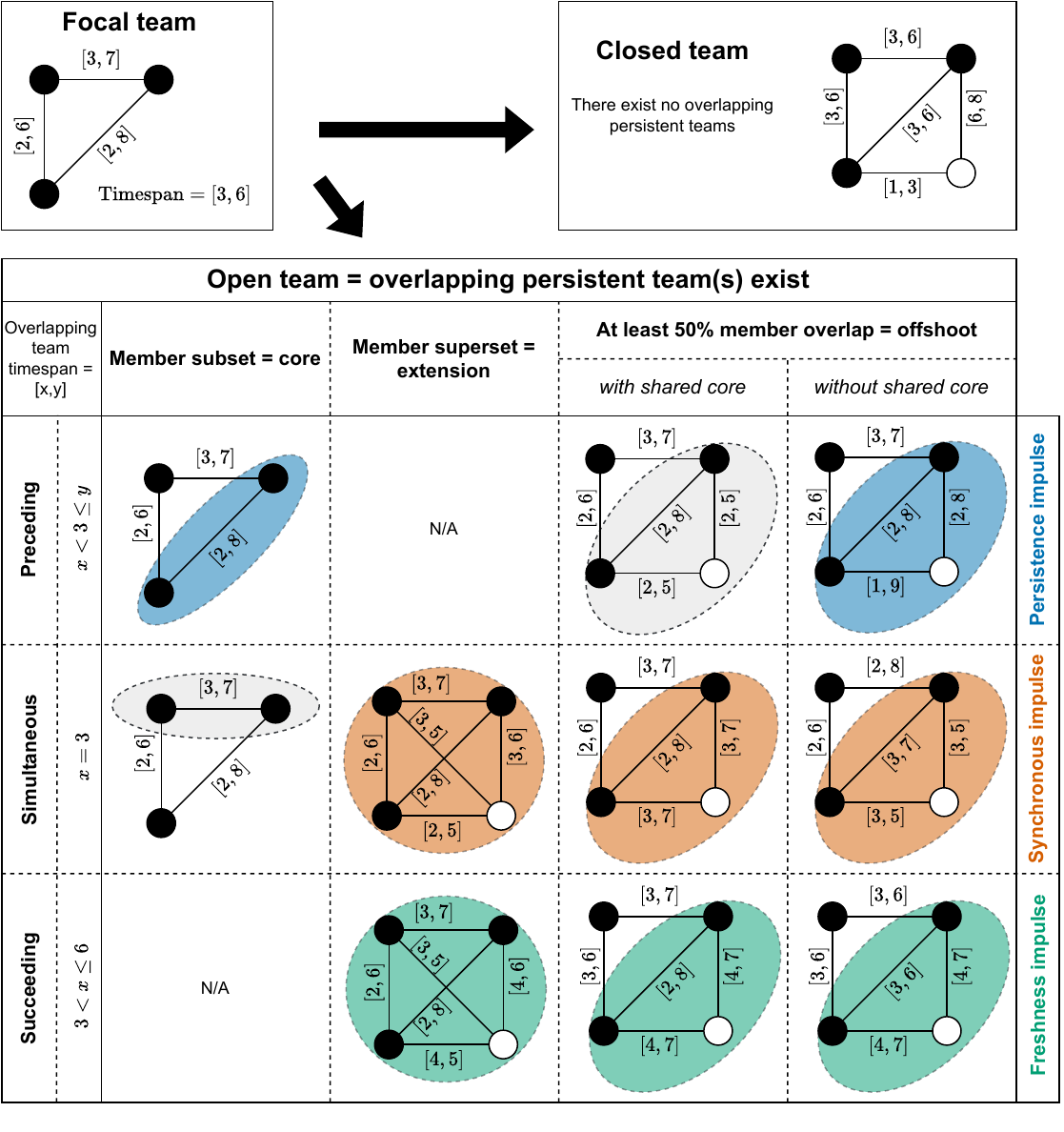}
    \caption{\textbf{Closed vs. open teams and the persistence, synchronous, and freshness impulses.} Overview of the various forms of team openness (columns) that are possible and how they can be categorized into three types of impulses (colors) for the focal team, based on their relative timing (rows).
    See Supplementary section Materials and Methods for explicit definitions of the different types of overlapping teams.}
    \label{fig:4}
\end{figure}

We distinguish between \emph{closed} and \emph{open teams}. 
Closed teams do not engage in \emph{overlapping persistent collaborations}, while open teams do. 
There are several distinctions to be made, depending on composition and timing of the overlapping teams, as outlined in Figure~\ref{fig:4}. 

\begin{figure}[!t]
    \centering
    \begin{subfigure}[t]{0.412\linewidth}
        \centering
        \includegraphics[width=\linewidth]{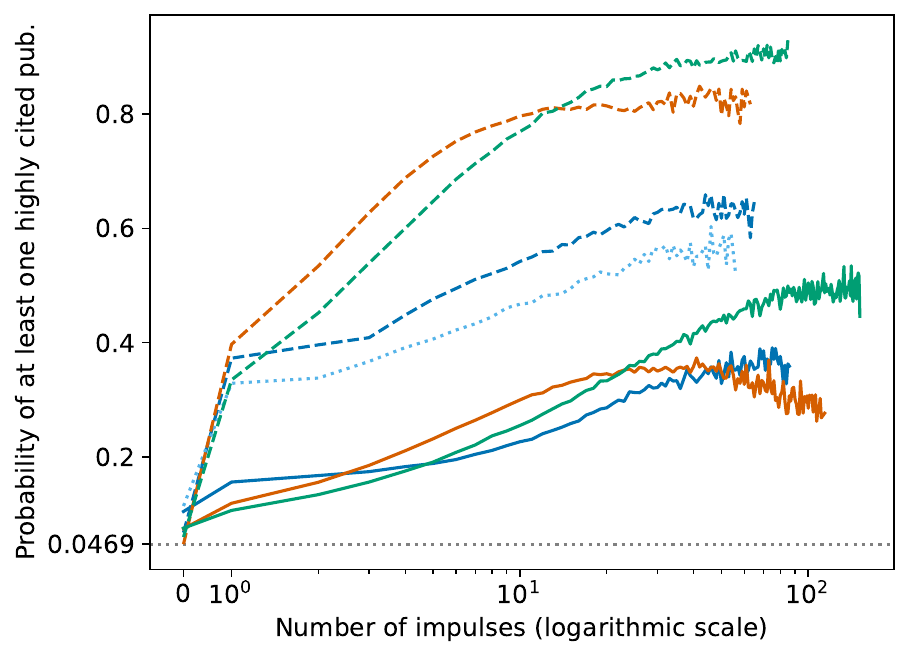}
        \caption{Probability that a team is top 1\% cited}
        \label{fig:5a}
    \end{subfigure}
    ~
    \begin{subfigure}[t]{0.56\linewidth}
        \centering
        \includegraphics[width=\linewidth]{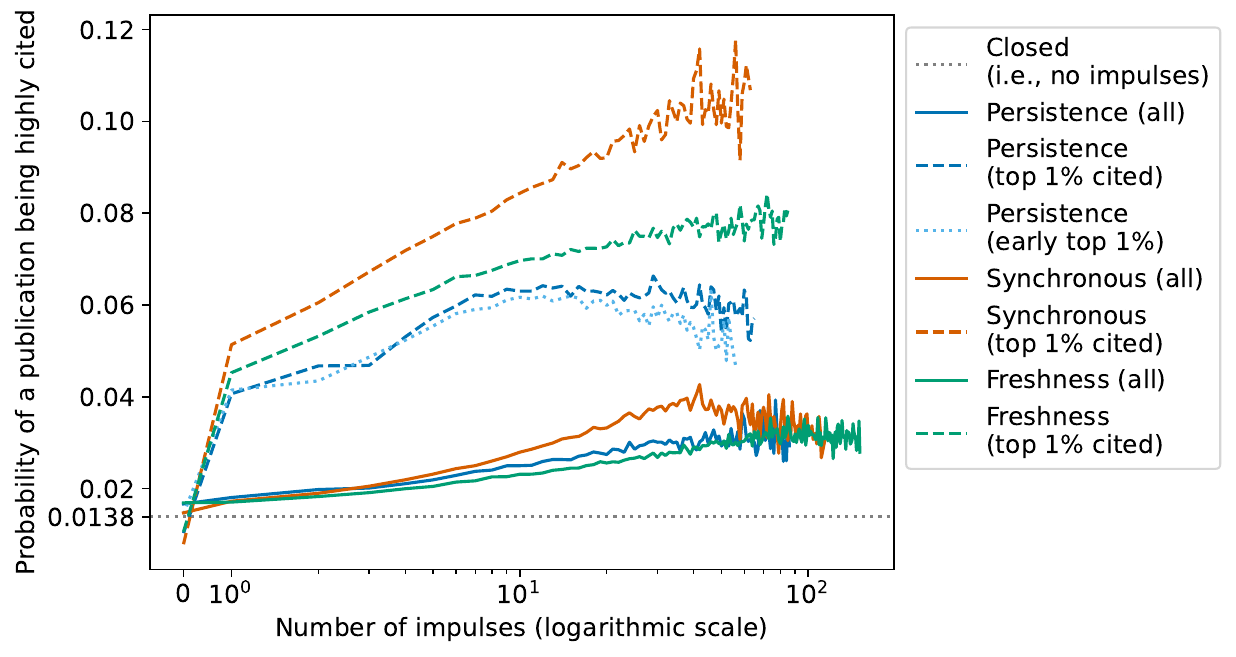}
        \caption{Probability that any one publication is top 1\% cited}
        \label{fig:5b}
    \end{subfigure}
    \begin{subfigure}[t]{0.474\linewidth}
        \centering
        \includegraphics[width=\linewidth]{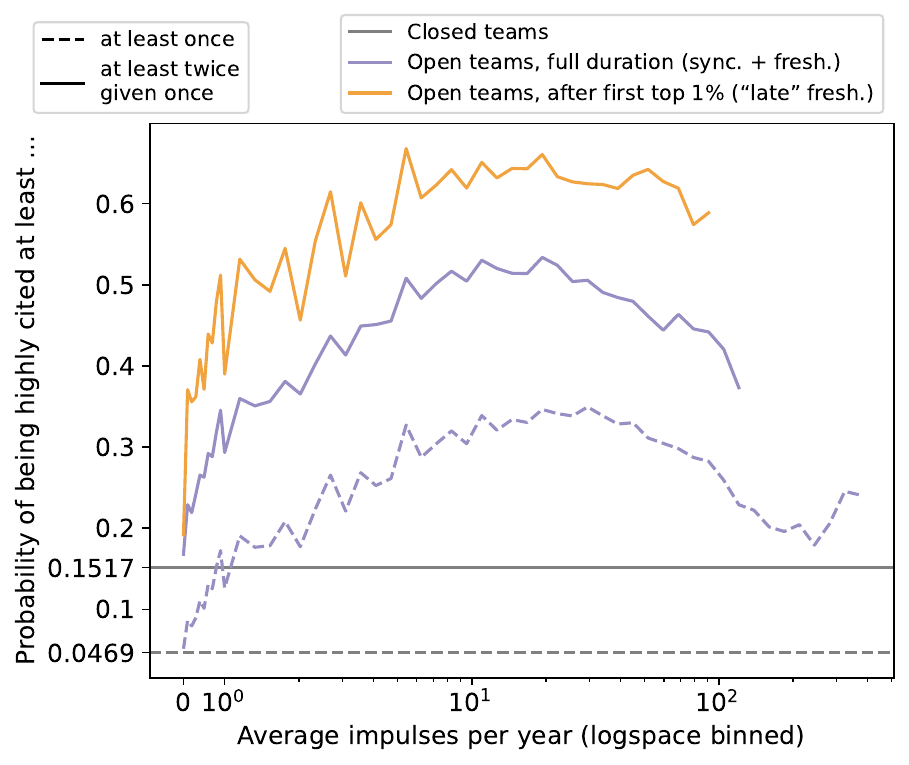}
        \caption{Probability of (additional) success}
        \label{fig:5c}
    \end{subfigure}
    ~
    \begin{subfigure}[t]{0.49\linewidth}
        \centering
        \includegraphics[width=\linewidth]{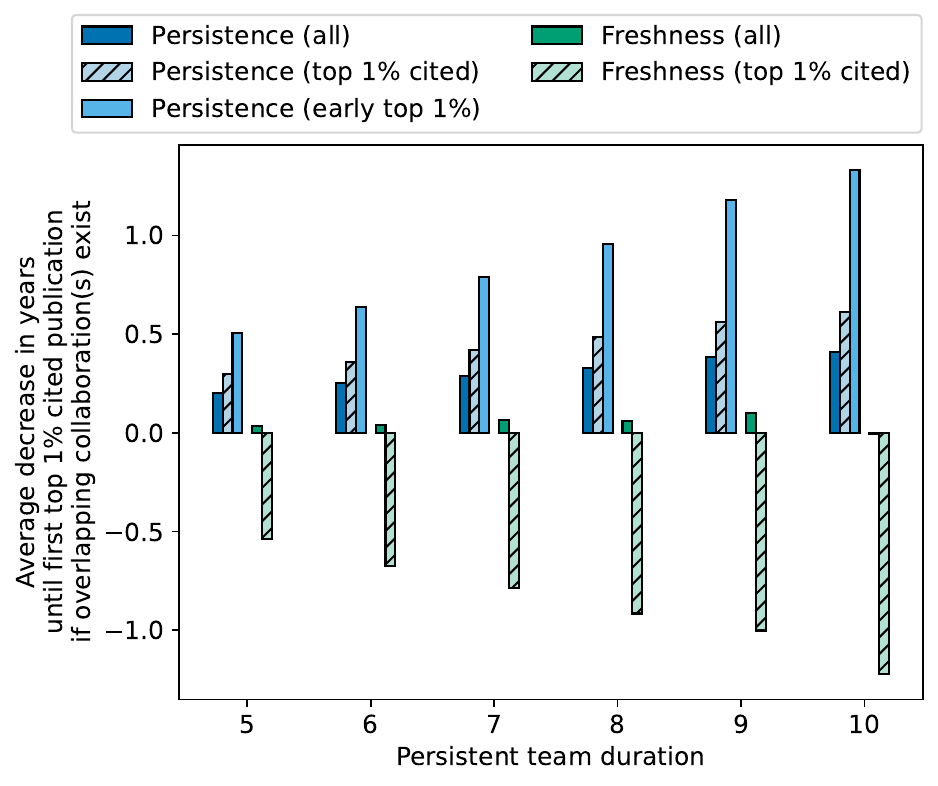}
        \caption{Decrease in year of first success}
        \label{fig:5d}
    \end{subfigure}
    
    \caption{\textbf{Impact of persistence, synchronous, and freshness impulses on citation success.} Panel (a) indicates the odds of a team to have at least one highly cited (top 1\%) publication given a certain number of impulses. Panel (b) indicates the probability for any given publication of a team with a certain number of impulses to be highly cited (top 1\%). Panel (c) shows the probability of having at least one (or at least two) top 1\% cited publication(s) based on the average number of new impulses per year. Finally, panel (d) demonstrates, for a range of team durations, the impact of persistence and freshness impulses on how early, on average, the first top 1\% publication is published by successful teams, regardless of the number of impulses. In panels (a), (b), and (c), the horizontal lines indicating the success of closed teams (receiving no impulses throughout their lifespan) are included for easy visual comparison.}
    \label{fig:5}
\end{figure}

If we consider composition (the columns of the table in Figure~\ref{fig:4}), overlapping persistent collaborations may take two forms: 1) they may be a collaboration of (at least half of) the focal team's members with new authors, with a (partial) timespan overlap; or 2) they may be a collaboration that fully encompasses and exceeds the timespan of the focal team by a subset of its members, i.e., a core team.
Time-wise (the rows of the table in Figure~\ref{fig:4}), these collaborations may commence preceding, simultaneous with, or succeeding the focal team. When at the start of a focal team, a subset of its members is already engaged in another persistent collaboration, this gives the focal team a \emph{persistence impulse} that may reduce the coordination costs in the new team. In similar vein, when during their persistent collaboration some team members work together with others, the focal team can benefit from \emph{freshness impulses}. Finally, a focal team can have an overlap with one or more teams that have their first recorded publication in the same year as the focal team, which we call \emph{synchronous impulses}.  
As the relative timing of synchronous impulses with respect to the focal team is unknown, synchronous impulses can, although clearly signaling openness, not unambiguously be categorized as freshness or persistence impulses. 
Hence, we exclude synchronous impulse results when an interpretation of relative timing is required, such as in Figure~\ref{fig:5d}.
We provide precise definitions and interpretations for each type of overlapping team shown in Figure~\ref{fig:4} in the Supplementary section Materials and Methods.
With this categorization we can investigate how openness through team overlap can impact the success of a team.

Figures~\ref{fig:5a} and~\ref{fig:5b} show that open teams are far more likely to have at least one highly cited publication, and are more likely to produce highly cited works in general, than closed teams. Being open leads to better science.
The existence of even a single overlap can significantly improve the odds of success, especially when those overlapping teams are themselves highly cited (i.e., top 1\% impulses). Teams that are more open and benefit from more impulses, tend to experience greater success. This generally holds for all types of impulses, persistence, synchronous, and freshness. But, just as we saw with diversity in team composition, after a certain point coordination costs outweigh the positive effect of the impulses. 
The synchronous impulse trends in Figures~\ref{fig:5a} and~\ref{fig:5b}, which include only one  year of impulses, together with Figure~\ref{fig:5c}, suggest engaging in too many new collaborations at a time becomes detrimental for success. 

How do teams use these impulses? In line with our previous findings we find that that teams benefit from freshness impulses through collaborations to remain successful through their lifespan (Figure~\ref{fig:5c}). Interestingly, we find that freshness impulses play only a minor role in achieving early success (see Figure~\ref{fig:5d}). In fact, the presence of highly cited freshness impulses is associated with, on average, later success (though Figure~\ref{fig:S4d} suggests that this is not an issue for top 10\% citation success). Thus, a team typically has success early on and can remain successful if its members engage in collaborations that spark freshness once the focal team is well established (but not early on). 
Teams can also use openness to source persistence impulses. Our findings show that in fact, early success of focal teams is closely linked to receiving these persistence impulses. This is especially the case when the team providing the persistence impulse was already successful before the focal team formed (\emph{early top 1\% persistence}). 

Thus, overlapping collaborations that bring experience (of success) with them are crucially beneficial to reduce early coordination costs and steer a new team towards success. Taking into account complex patterns of team overlap therefore allows for a nuanced understanding of how freshness and persistence both drive academic success, but at different stages in a team process. 

\section*{Conclusion}

Using a novel network-driven approach, we identified scientific teams that capture the fluid and often interdisciplinary nature of modern science. 
We showed how teams are highly prevalent in global scientific output, and clearly overrepresented among highly cited works.
Moreover, we found that a team's freshness is a greater driver of success than its persistence.
However, we observed that when team members displayed persistence through collaborations (with outside authors) before the team's formation, this can drastically improve the team's ability to produce high-impact work early. 
Additionally, a team's openness to new persistent collaborations during its lifetime, allows the team to ``stay fresh'' and achieve continued success. 

In conclusion, a team's success, in terms of citations, follows primarily from the knowledge recombination of its own freshness and the freshness impulses it receives through outside collaborations, as well as team diversity. Persistent collaboration instead facilitates the success of such knowledge recombination(s) by reducing some of the early coordination costs of these fresh and diverse teams.

In light of these findings, it is interesting to revisit our observation that the academic powerhouses USA and China both show relatively low rates of persistent collaboration. They sustain their impressive levels of high impact scientific knowledge creation~\cite{pisani2024china} by steering clear of persistent collaborations. Other countries may do well to follow in this pathway and gear their academic policies and grant schemes towards supporting exciting research of fresh teams, that benefit from some previous collaboration in the core team, rather than expecting the continuation of early success.  
Finally, for scholars struggling to make impact, our work suggests that, when a collaboration does not pay off in the first few years, it is advisable to seek new collaborations. But it is advised to do so together with (some) existing collaborators.


\bibliography{persistentteams}
\bibliographystyle{sciencemag}


\section*{Acknowledgments}
The authors thank colleagues at the Centre for Science and Technology Studies (CWTS) at Leiden University for their hospitality and for generously providing the first author with access to the CWTS database during his time as a Ph.D. candidate at Leiden University.
\paragraph*{Funding:}
There are no funding sources to declare.
\paragraph*{Author contributions:}
H.D.B: Conceptualization, Data curation, Formal analysis, Investigation, Methodology, Software, Visualization, Writing -- original draft, Writing -- review \& editing. \\
E.M.H.: Conceptualization, Writing -- review \& editing. \\ 
N.P.: Conceptualization, Writing -- review \& editing. \\
F.W.T.: Conceptualization, Supervision, Writing -- review \& editing
\paragraph*{Competing interests:}
There are no competing interests to declare.
\paragraph*{Data and materials availability:}
The bibliographic dataset used in this work, is based on Web of Science (WoS). Downstream aggregated data on teams and their success, from which the results and figures in this paper are determined, are available on Zenodo~\cite{boekhout_2025_15967685}.


\subsection*{Supplementary materials}
Materials and Methods\\
Supplementary Text\\
Figs. S1 to S4\\
Table S1\\
References \textit{(21-\arabic{enumiv})}\\


\newpage


\renewcommand{\thefigure}{S\arabic{figure}}
\renewcommand{\thetable}{S\arabic{table}}
\renewcommand{\theequation}{S\arabic{equation}}
\renewcommand{\thepage}{S\arabic{page}}
\setcounter{figure}{0}
\setcounter{table}{0}
\setcounter{equation}{0}
\setcounter{page}{1} 


\begin{center}
\section*{Supplementary Materials for\\ \scititle}

Hanjo~D.~Boekhout$^{\ast}$,
Eelke~M.~Heemskerk,
Niccol\`{o}~Pisani,
Frank~W.~Takes\\ 
\small$^\ast$Corresponding author. Email:h.d.boekhout@liacs.leidenuniv.nl\\
\end{center}

\subsubsection*{This PDF file includes:}
Materials and Methods\\
Supplementary Text\\
Figures S1 to S4\\
Table S1

\newpage


\subsection*{Materials and Methods}

In this section we first describe our dataset and how we determine successful publications.
Then, we discuss the proposed network-driven methodology to identify persistent teams and their publications from this dataset, as well as an explanation of how we, based on the overlap in network structure, categorize different types of team openness as persistence or freshness impulses. 

\subsubsection*{Dataset}

The results presented in this work are based on Clarivate's Web of Science database (WoS). 
Specifically, the data we used was collected from the in-house version of WoS at the Centre for Science and Technology Studies (CWTS) at Leiden University from April 2023. 
The in-house version of WoS has been enriched by CWTS in a number of ways relevant to this research. 
Specifically, CWTS 1) performs its own citation matching (i.e., matching of cited references to the publications they refer to); 2) pays particular attention to the allocation of publications to universities and organizations in a consistent and accurate manner~\cite{waltman2012leiden}; and 3) performs geocoding of the addresses listed in publications and disambiguated authors~\cite{caron2014large}. 
We have further enriched this dataset by allocating each geocoded address and affiliation to a scientific city~\cite{boekhout2022evolution}.

We consider publications published in 2008--2020 that are categorized as either Article, Review, Letter or Proceeding Paper. 
Publications with missing author-affiliation linkages are excluded, as well as publications for which one (or more) of its listed affiliations has neither an associated geolocation nor organization (including universities). 
This leaves us with a total of 25.2~million publications (86.7\%), covering 198 countries and nearly 16.5~thousand scientific cities with 31.8~million unique authors. 
Highly cited publications (top 10\% and top 1\%) are determined per publication year and scientific field (WoS subject category) to prevent fields with high average citations from dominating the highly cited publications. 
We determine the highly cited publications based on citations received within three years of publication. 
Table~\ref{tab:data-stats} provides statistics on the prevalence of the document types.

\begin{table} 
	\centering
	\caption{\textbf{Document type prevalence.}}
	\label{tab:data-stats} 
	\begin{tabular}{l|rrr} 
		\hline
        & \multicolumn{3}{c}{Percentage of publications per type} \\
		Document type & All & Top 10\% cited & Top 1\% cited\\
		\hline
		Article           & 78.35 & 83.02 & 72.76 \\
		Review            &  4.80 & 12.87 & 23.80 \\
		Letter            &  1.76 &  0.31 &  0.38 \\
        Proceedings Paper & 15.09 &  3.79 &  3.06 \\
		\hline
	\end{tabular}
\end{table}

\subsubsection*{Identifying persistent teams and their publications}

\begin{figure}[!t]
    \centering
    \includegraphics[width=\linewidth]{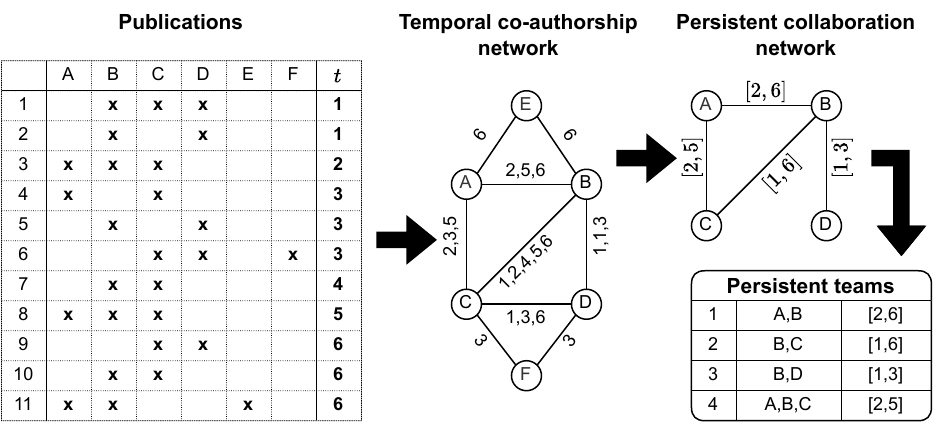}
    \caption{\textbf{Example of the process for identifying persistent teams from publication data.} In the publication table, each row represents a publication while the columns A, B, $\ldots$, F refer to the authors and column $t$ indicates the year of publication.}
    \label{fig:pers-teams-process}
\end{figure}

The process of identifying persistent teams from publication data consists of three steps (as illustrated through an example in Figure~\ref{fig:pers-teams-process}: 1) constructing the temporal co-authorship network; 2) transforming this into a persistent collaboration network; and 3) extracting persistent teams through temporal maximal clique enumeration on the persistent collaboration network.

The \emph{temporal co-authorship network} constructed in the first step, consists of nodes, which represent disambiguated authors, and edges connecting these nodes representing the co-authorship(s) between authors.
The year(s) and frequency of co-authorship(s) between authors, are captured as edge attributes.
Multiple approaches to capturing this information can be taken, such as an edge for each co-authorship labeled with the year of publication or an edge for each year of publication labeled with the frequency of co-authorship in that year.
For simplicity, our example in Figure~\ref{fig:pers-teams-process} shows the timestamps, i.e., years, of each co-authored publication as a comma separated list. 

In the second step, we transform this temporal co-authorship network into a \emph{persistent collaboration network}. 
The resulting persistent collaboration network captures, for each pair of authors, which periods of years they are sufficiently persistent.
Note that, during the transformation, nodes remain unchanged, i.e., they continue to represent the same authors. 
We consider a pair's collaboration sufficiently persistent when they co-author at least three publications within a five year period. 
As illustrated by our example in Figure~\ref{fig:pers-teams-process}, the actual period of persistent collaboration can be less than five years, as it is bounded by the actual co-authorship years, but can also exceed five years as long as the three co-authorship minimum is satisfied for each five year stretch within, such as for author pair (B,C) in our example.
Note that in the example, the co-authorship between author pair (C,D) is not sufficiently persistent as their three publications span over a six year period.
Furthermore, note that it is possible for an author pair to have multiple disconnected periods of persistent collaboration. 
We acknowledge that using one global definition of persistent collaboration does not account for differences in, for example, the rate of publication between fields. 
Consequently, for fields with low rates of publication, persistent collaboration may be underestimated.

Finally, in the third step we extract persistent teams from the persistent collaboration network.
To do so, we employ the temporal maximal clique enumeration algorithm recently introduced by Boekhout \& Takes~\cite{boekhout2024fastmaximalcliqueenumeration}. 
The method enumerates so-called ($\delta, \gamma$)-maximal cliques. 
A \emph{($\delta, \gamma$)-clique} is a set of nodes and a timespan, such that the nodes are fully connected with a minimum weight of $\gamma$ during every $\delta$ length period in the timespan.
A clique is considered \emph{maximal} when extending either the node set or timespan, without altering the other, would no longer qualify it as a ($\delta, \gamma$)-clique.
Here, we use $\delta = 1$ and $\gamma = 1$ and consider, for each pair of authors in the persistent collaboration network, edges to exist for each year in their persistent collaboration periods. Thus, in our example in Figure~\ref{fig:pers-teams-process}, node pair (A,B) has edges for years 2, 3, 4, 5, and 6. Notice how this enables us to find persistent team {A,B,C} with timespan $[2,5]$ despite author pair (A,B) only publishing two papers in this timespan. 
Had we applied the temporal maximal clique enumeration directly on the temporal co-authorship network with $\delta = 5$ and $\gamma = 3$, then this persistent team would not be found despite each pair having clearly overlapping persistent collaboration periods.
Thus, by first transforming the co-authorship network into a persistent collaboration network, we are able to find persistent teams that more accurately capture the dynamic nature of scientific teams.

For each persistent team, we next determine the set of publications to which it has contributed. We consider a publication to be associated with a team when at least half, but no less than two, of the members of the team are included among its authors, and the publication was published during the team's persistent timespan.
We use this approach to not unfairly assign a publication to large teams where the majority was uninvolved.

If due to disconnected periods of persistent collaboration between team members, a team also has disconnected persistent timespans, that team is considered to have a duration that spans all disconnected timespans. We use this approach to prevent incorrect assumptions of team freshness for later periods. However, to prevent publication successes (and non-successes) during years that the team was not considered to exist from affecting the analysis, these publications are excluded for that team. Thus, only the publications from years during the disconnected timespans are considered in our analysis.

\subsubsection*{Overlapping persistent teams}

In this section we elaborate on the different ways in which persistent teams may overlap, as visualized in Figure 4. We explain, for each type of member overlap and relative timing of this overlap (i.e., for each cell in Figure 4), what type of impulse (if any) such a type of openness, represents. 
Throughout, we consider our focal team ($T_f = (M_f,[x_f,y_f])$) to have member set $M_f$ and timespan $[x_f, y_f]$.

Starting with the leftmost column, we consider \emph{core} overlapping teams ($T_c = (M_c,[x_c,y_c])$) whose member sets are a subset of that of the focal team, i.e., $M_c \subset M_f$.
Since every pair of authors in a team must be connected for every year in its duration, we know that $x_c \leq x_f$ and $y_f \leq y_c$. 
Because $x_c \ngtr x_f$, succeeding timing (i.e., where the core team starts after the focal team) is not possible and we show this as ``N/A''.
When $x_c = x_f$, i.e., for simultaneous timing, we know that any first year publication associated with the core team is most likely also associated with the focal team and therefore provides no impulse. For this reason, the core team is shown in light gray in Figure~\ref{fig:4} and not counted towards the synchronous impulses.
Consequently, the only relevant relative timing for core teams is the preceding timing, i.e., $x_c < x_f \leq y_c$. 
Here, the core team may represent, for example, a Ph.D.'s supervisors or the core members of a research group. Such core teams bring their collaborative experience with them into the new collaboration and thus give a persistence impulse.

The second column considers \emph{extension} overlapping teams ($T_e = (M_e,[x_e,y_e])$). Extension teams are defined by their member sets that are a superset of the focal team's members, i.e., $M_f \subset M_e$.
As extension teams are essentially the inverse of core teams, we know that $x_f \leq x_e$ and $y_e \leq y_f$. 
As such, $x_e \nless x_f$ and preceding extension teams are impossible (``N/A'').
For simultaneous timing and succeeding timing (the middle and bottom rows), the added author(s) ensure that the extension teams introduce at least some new publications and therefore supply, respectively, synchronous and freshness impulses. 
Extension teams may represent shorter duration collaborations of some core members with junior or temporary group members.

The third and fourth columns consider \emph{offshoot} overlapping teams ($T_o = (M_o,[x_o,y_o])$). The member sets of offshoot teams have at least a 50\% overlap with the focal team, but are neither a subset or superset, i.e., $\frac{1}{2}\max(|M_o|,|M_f|) \leq |M_o \cap M_f|$ and $M_o \not\subset M_f \not\subset M_o$.
As the focal and offshoot teams can have a shared core team, we distinguish between the cases that do (third column) and those that do not (fourth/last column).
We make this distinction because when a preceding offshoot has a shared core team, both the offshoot and the core team provide essentially one persistence impulse. Therefore, we choose not to count the preceding offshoots with shared cores as persistence impulses and they are shown in light gray in Figure~\ref{fig:4}. We do count preceding offshoots without a shared core as persistence impulses.
For both simultaneous and succeeding timing this double counting is not an issue and both the offshoots with and without shared core are considered to provide, respectively, synchronous and freshness impulses.


\subsection*{Supplementary Text}
\label{suptext}

As a robustness check, we compare the results presented in the main text on the relationship between top 1\% citation success and team freshness and openness, with those for top 10\% citation success. 
Note that the most important differences in conclusions were already highlighted in the main text.

\begin{figure}[t]
    \centering
    \begin{subfigure}[t]{0.48\linewidth}
        \centering
        \includegraphics[width=\linewidth]{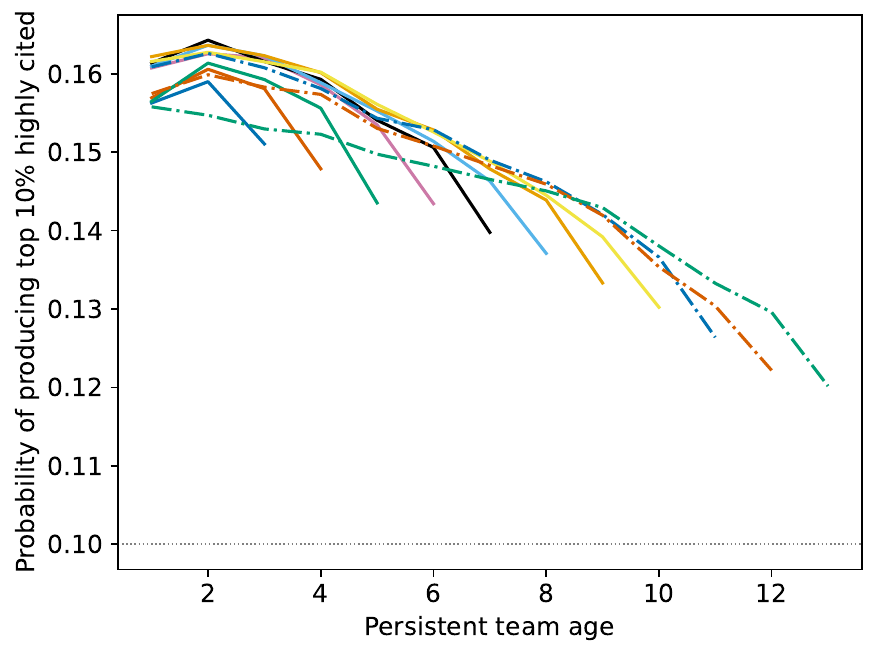}
        \caption{Success probability over time}
        \label{fig:s2a}
    \end{subfigure}
    ~
    \begin{subfigure}[t]{0.48\linewidth}
        \centering
        \includegraphics[width=\linewidth]{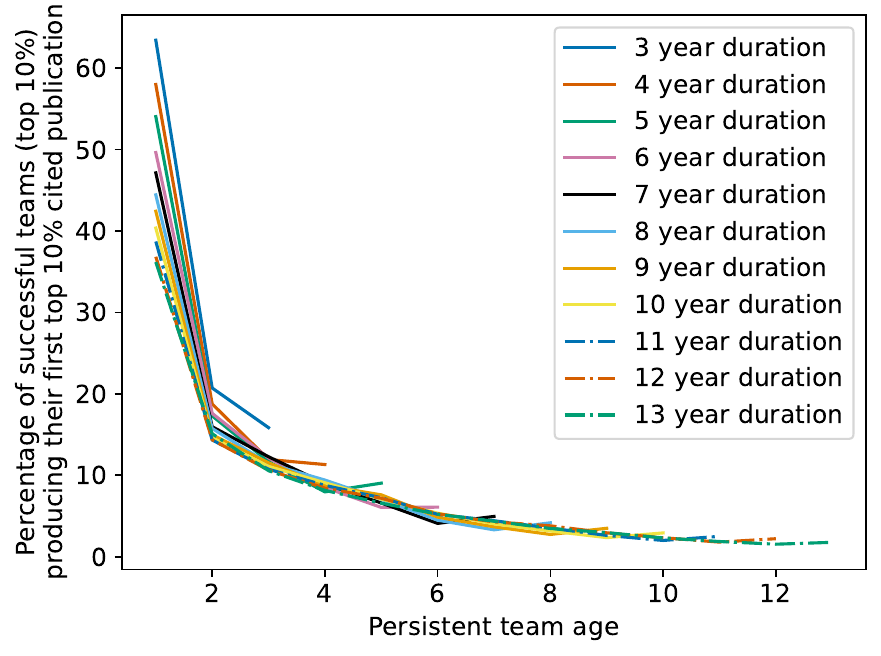}
        \caption{Year of first success}
        \label{fig:s2b}
    \end{subfigure}
    \caption{\textbf{Team freshness vs. top 10\% citation success.} Panel (a) shows the probability that publications, produced by teams of a given age, become highly cited. Panel (b) indicates the percentage of highly cited teams that have their first highly cited publication at a given team age.}
    \label{fig:s2}
\end{figure}

\begin{figure}[t]
    \centering
    \begin{subfigure}[t]{0.49\linewidth}
        \centering
        \includegraphics[width=\linewidth]{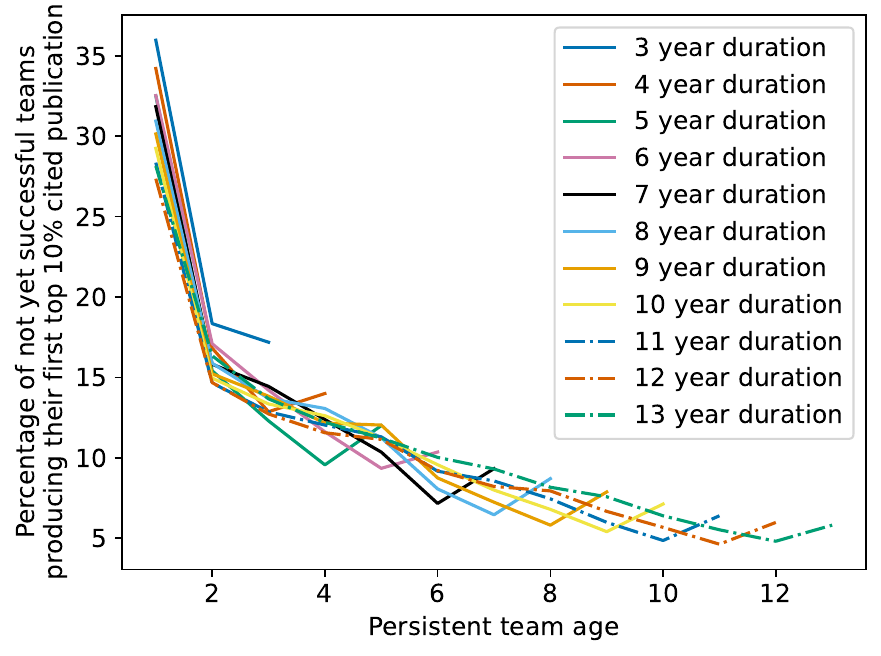}
        \caption{Success probability over time}
        \label{fig:s2c}
    \end{subfigure}
    ~
    \begin{subfigure}[t]{0.49\linewidth}
        \centering
        \includegraphics[width=\linewidth]{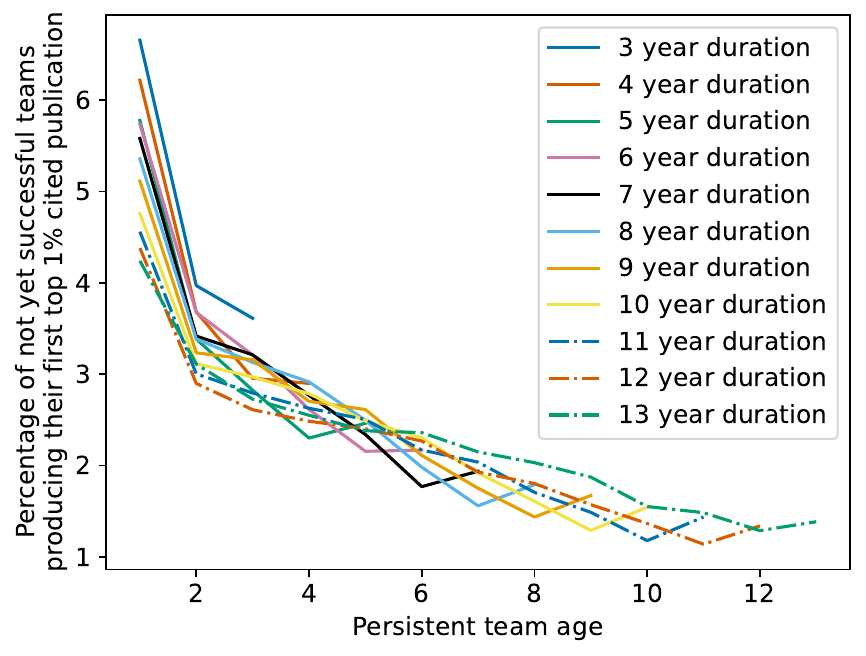}
        \caption{Year of first success}
        \label{fig:s2d}
    \end{subfigure}
    \caption{\textbf{Team freshness vs. citation success.} Panels indicate for each given team age, the percentage of not yet successful teams that have their first highly cited publication that year. Panel (a) shows this for top 10\% and panel (b) for top 1\% citation success.}
    \label{fig:s2addendum}
\end{figure}

\begin{figure}[!t]
    \centering
    \begin{subfigure}[t]{0.412\linewidth}
        \centering
        \includegraphics[width=\linewidth]{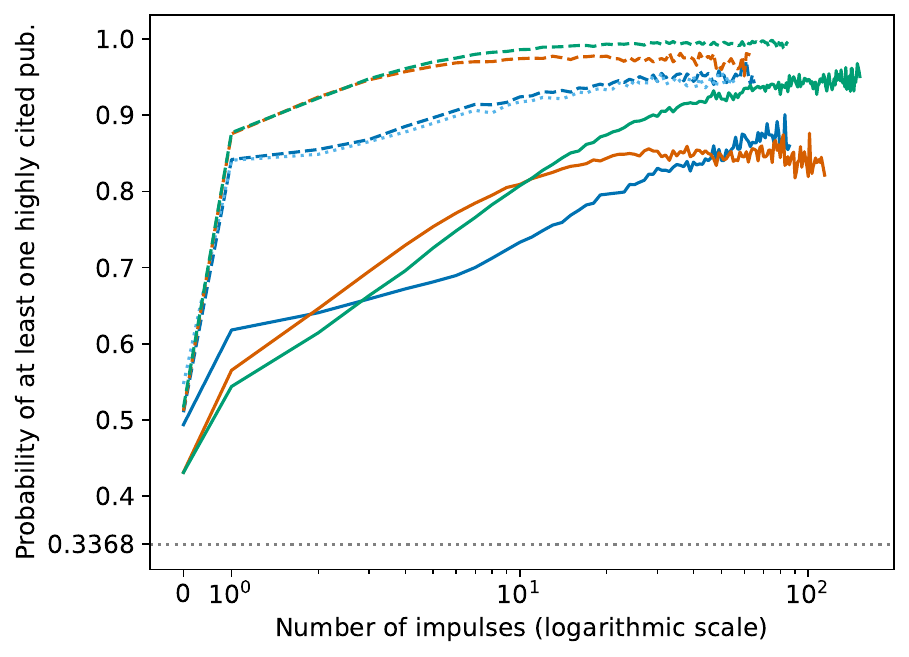}
        \caption{Probability that a team is top 10\% cited}
        \label{fig:S4a}
    \end{subfigure}
    ~
    \begin{subfigure}[t]{0.56\linewidth}
        \centering
        \includegraphics[width=\linewidth]{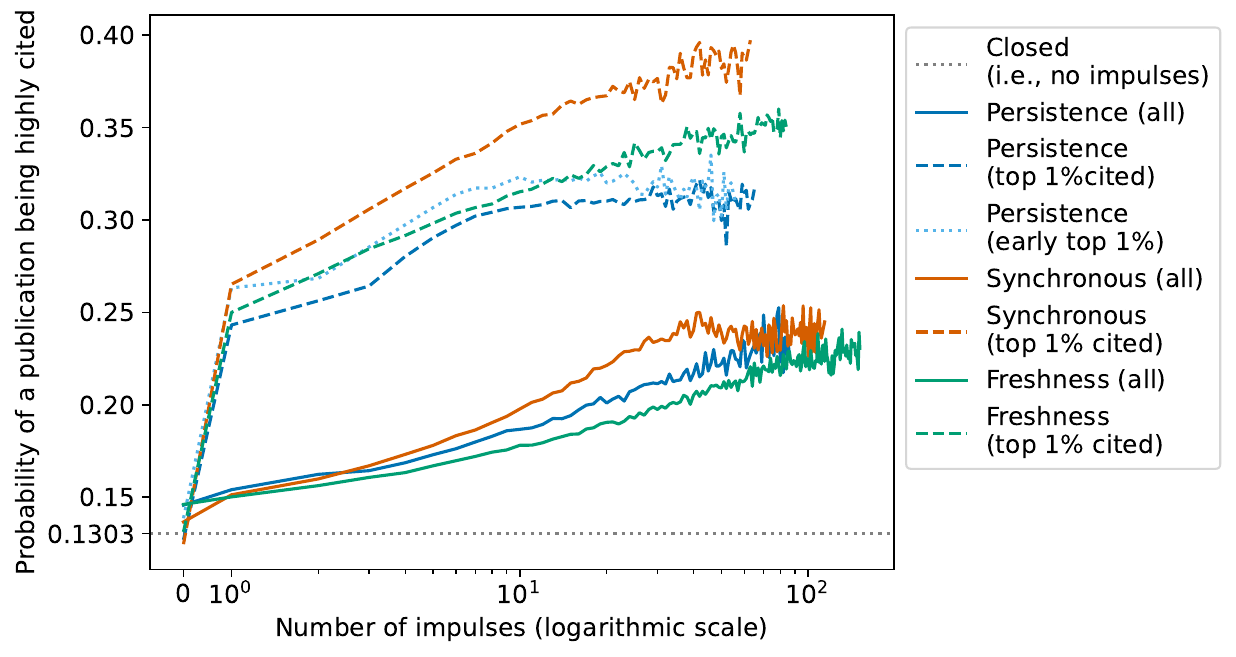}
        \caption{Probability that any one publication is top 10\% cited}
        \label{fig:S4b}
    \end{subfigure}
    \begin{subfigure}[t]{0.474\linewidth}
        \centering
    \includegraphics[width=\linewidth]{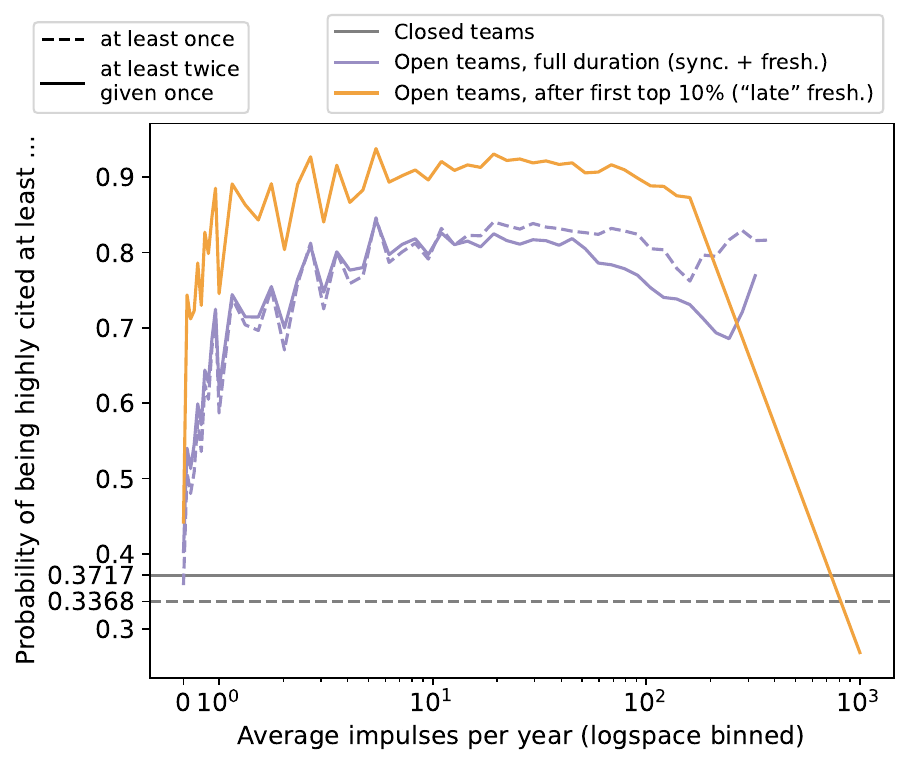}
        \caption{Probability of (additional) success}
        \label{fig:S4c}
    \end{subfigure}
    ~
    \begin{subfigure}[t]{0.49\linewidth}
        \centering
        \includegraphics[width=\linewidth]{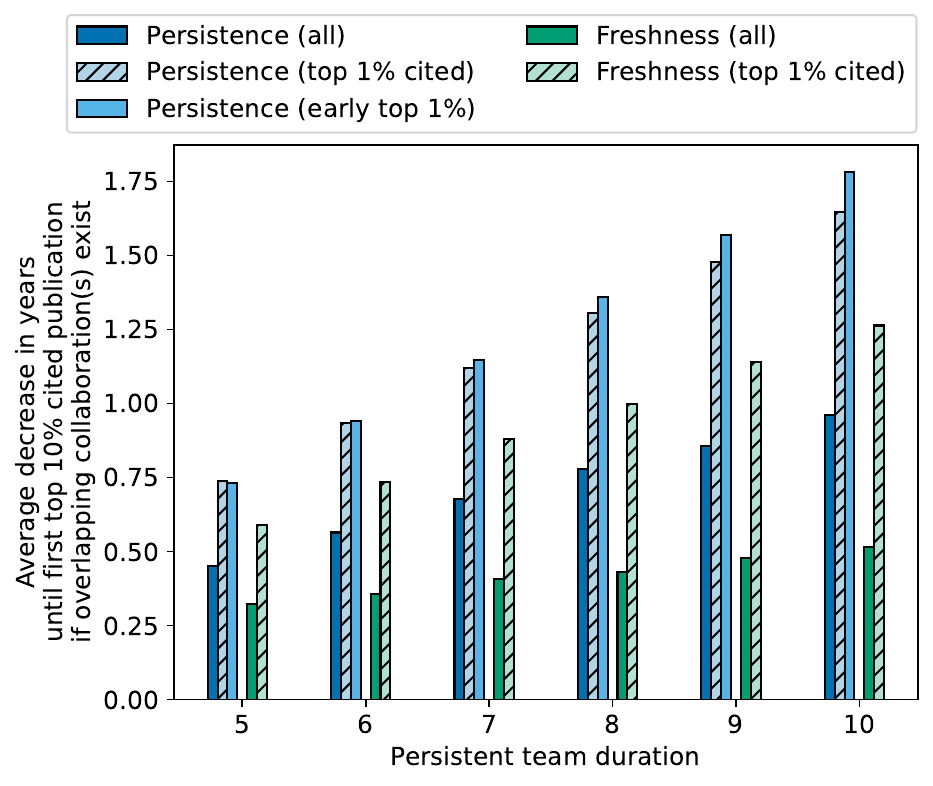}
        \caption{Decrease in year of first success}
        \label{fig:S4d}
    \end{subfigure}
    
    \caption{\textbf{Impact of persistence, synchronous, and freshness impulses on citation success.} Panel (a) indicates the odds of a team to have at least one highly cited (top 10\%) publication given a certain number of impulses. Panel (b) indicates the probability for any given publication of a team with a certain number of impulses to be highly cited (top 10\%). Panel (c) shows the probability of having at least one (or at least two) top 10\% cited publication based on the average number of new impulses per year. Finally, panel (d) demonstrates, for a range of team durations, the impact of persistence and freshness impulses on how early, on average, the first top 10\% publication is published by successful teams, regardless of the number of impulses. In panels (a), (b), and (c), the horizontal lines indicating the success of closed teams (receiving no impulses throughout their lifespan) are included for easy visual comparison.}
    \label{fig:S4}
\end{figure}

With regards to team freshness, Figures~\ref{fig:s2} and~\ref{fig:s2addendum} show the same trends for top 10\% success as observed for top 1\% success. In other words, success tends to come early and the odds of becoming successful decrease over time.
Similarly with regards to team openness, Figures~\ref{fig:S4a} and~\ref{fig:S4b} show similar trends for top 10\% success as observed for top 1\% success. Thus, open teams are also more successful for top 10\% success.
A minor difference can be observed in Figure~\ref{fig:S4c} which looks at the impact of yearly freshness impulses on (continued) success. We observe the same general curvature for each line, but the odds of first success are much closer to the odds of continued success. However, this difference speaks more to the fact that each publication is by definition ten times more likely to become top 10\% highly cited than top 1\%.
Furthermore, the probabilities for first and second success being closer to each other have no impact on the relevant conclusion. Thus, for top 10\% success as well, engaging in too many new collaborations at a time becomes detrimental.

As mentioned in the main text, the only real difference we found between top 10\% success and top 1\% success, is in the relationship between  freshness impulses and early success. Whereas these impulses have a negative, or at best neutral, relation with top 1\% success, Figure~\ref{fig:S4d} demonstrates a positive relationship between the (highly cited) freshness impulses and early top 10\% success. 
Given the very high probability of becoming successful for open teams in general observed in Figure~\ref{fig:S4a}, it is reasonable to assume that this positive relationship is more a consequence of the tendency for early success than a consequence of the freshness impulses.
Notably, the persistence impulses maintain a larger positive effect on early success than the freshness impulses.

\newpage


\end{document}